\documentclass{ws-procs9x6}

\begin{document}

\title{POPULATION OF NEUTRON UNBOUND STATES VIA TWO-PROTON KNOCKOUT REACTIONS}

\author{N. FRANK,\footnote{Present address: Department of Physics, Concordia College, Moorhead, MN 56562} T. BAUMANN, D. BAZIN, A. GADE, J.-L. LECOUEY,\footnote{Present address: LPC, IN2P3, 14050 Caen, France} W.A. PETERS, H. SCHEIT\footnote{Present address: Nishina Center, RIKEN, Wako, Saitama 351-0198, Japan}, A. SCHILLER, and M. THOENNESSEN}

\address{National Superconducting Cyclotron Laboratory, Department of Physics \& Astronomy, \\ Michigan State University, East Lansing, MI 48824}

\author{J. BROWN}

\address{Department of Physics, Wabash College, Crawfordsville, IN 47933}

\author{P.A. DEYOUNG}

\address{Department of Physics, Hope College, Holland, MI 49423}

\author{J.E. FINCK}

\address{Department of Physics, Central Michigan University, Mt.\ Pleasant, MI 48859}

\author{J. HINNEFELD}

\address{Department of Physics \& Astronomy, IUSB, South Bend, IN 46634}

\author{R. HOWES}

\address{Department of Physics, Marquette University, Milwaukee, WI 53201}

\author{B. LUTHER}

\address{Department of Physics, Concordia College, Moorhead, MN 56562}

\begin{abstract}
The two-proton knockout reaction $^9$Be($^{26}$Ne,$^x$O2p) was used to explore excited unbound states of $^{23}$O and $^{24}$O. In $^{23}$O a state at an excitation energy of 2.79(13) MeV was observed. There was no conclusive evidence for the population of excited states in $^{24}$O.
\end{abstract}


\bodymatter

\section{Introduction}
Two-proton knockout reactions using intermediate-energy heavy-ion beams have been successfully used to explore the structure of neutron-rich nuclei \cite{Baz03,War03}. It has been shown that these reactions can be considered direct reactions. Spectroscopic factors can then be extracted from the longitudinal momentum distribution of the residual fragment. Spectroscopic factors of excited states can be extracted by $\gamma$-ray coincidence measurements.

Nuclei close to the dripline have very few if any bound excited states. The structure of these nuclei can be explored with neutron-decay spectroscopy \cite{Fra07a}. The evolution of the shell structure along the oxygen isotopes is of particular interest \cite{Ots01}. The two heaviest particle-bound oxygen isotopes $^{23}$O and $^{24}$O do not have any bound excited states \cite{Sta04}.

The MoNA collaboration has recently used the two-proton knockout reaction from $^{26}$Ne to populate excited states in $^{24}$O and $^{23}$O \cite{Fra07a,Sch07}.

\section{Experiment}

The experiment was performed at the National Superconducting Cyclotron Laboratory at Michigan State University. The secondary beam of 86~MeV/$u$ was produced from a primary $^{40}$Ar beam. The two-proton knockout reaction occurred in a 721 mg/cm$^2$ Be target located in front of a large-gap sweeper magnet \cite{Bir05}. The oxygen isotopes were detected and identified behind the magnet. Neutrons around zero degrees were measured in coincidence with the Modular Neutron Array (MoNA) \cite{Lut03,Bau05}. The details of the fragment production and separation as well as the experimental setup can be found in references \cite{Fra07a,Sch07,Fra06,Pet07}.

The decay energies of resonances were reconstructed by the invariant mass method from the measured oxygen and neutron energies and the opening angle between the oxygen and the neutron. The angle and energy of the fragments were ion-optically reconstructed using a novel method which takes into account the position at the target in the dispersive direction \cite{Fra07b}.

\section{Results}

\begin{figure}[t]
\begin{center}
\psfig{file=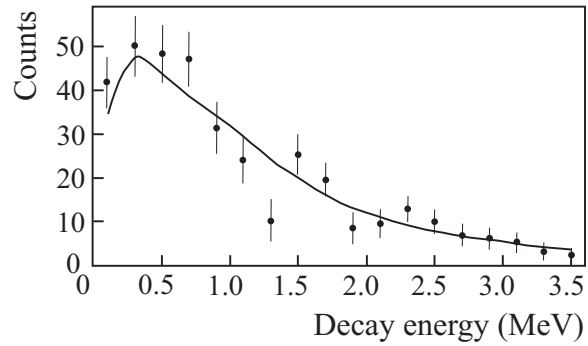,width=3.in}
\end{center}
\caption{Decay energy spectrum of $^{24}$O (data points). The solid line was calculated from a simulated thermal model.}
\label{fig1}
\end{figure}

\begin{figure}[b]
\begin{center}
\psfig{file=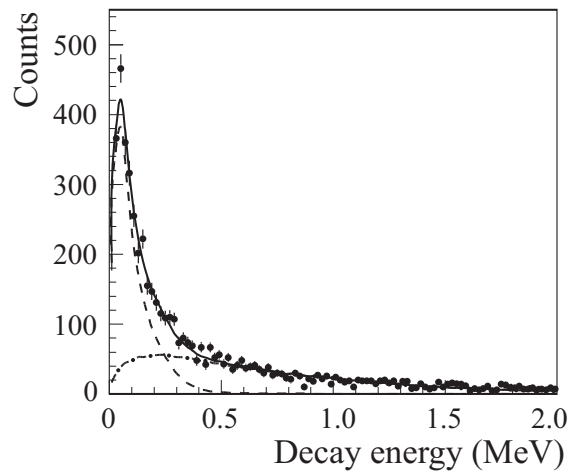,width=2.9in}
\end{center}
\caption{Decay-energy spectra of $^{23}$O$^*$ (data points). The solid line corresponds to the sum of the dashed (simulated resonant contribution) and dash-dotted (simulated thermal model) lines \protect \cite{Fra07a}.}
\label{fig2}
\end{figure}

Figure \ref{fig1} shows the reconstructed decay energy spectrum of excited $^{24}$O. No obvious resonance structure is apparent. The solid line corresponds to a background contribution simulated by a thermal distribution with a temperature of 1.7 MeV. 
In contrast, the decay energy spectrum of $^{23}$O shown in Figure \ref{fig2} shows a sharp resonance structure very close to threshold \cite{Fra07a,Sch07,Fra06}. The solid line corresponds to a fit with a resonance at 45(2) keV and a background contribution with a temperature of 0.7 MeV. With a neutron separation energy of 2.74(13) MeV \cite{Aud03} this corresponds to a state at an excitation energy of 2.79(13) MeV. The state is interpreted as the $5/2^+$ state due to the single particle hole in the $d_{5/2}$ configuration. The $3/2^+$ state recently observed for the first time in the reaction $^{22}$O(d,p)$^{23}$O \cite{Ele07} was not observed. 

The absence of excited states in $^{24}$O and the selective population of the $5/2^+$ state in $^{23}$O can be explained with the assumption of a direct knockout reaction and the structure of $^{26}$Ne. The knockout of the two valence protons ($\pi(0d_{5/2})^2$) from $^{26}$Ne will populate the 
particle-bound ground-state of $^{24}$O. The spectroscopic factor for the population of the unbound $2^+$ state from this reaction is predicted to be quite small \cite{Bro06}. It is not surprising that this state was not observed in the present reaction. 

\begin{figure}[b]
\begin{center}
\psfig{file=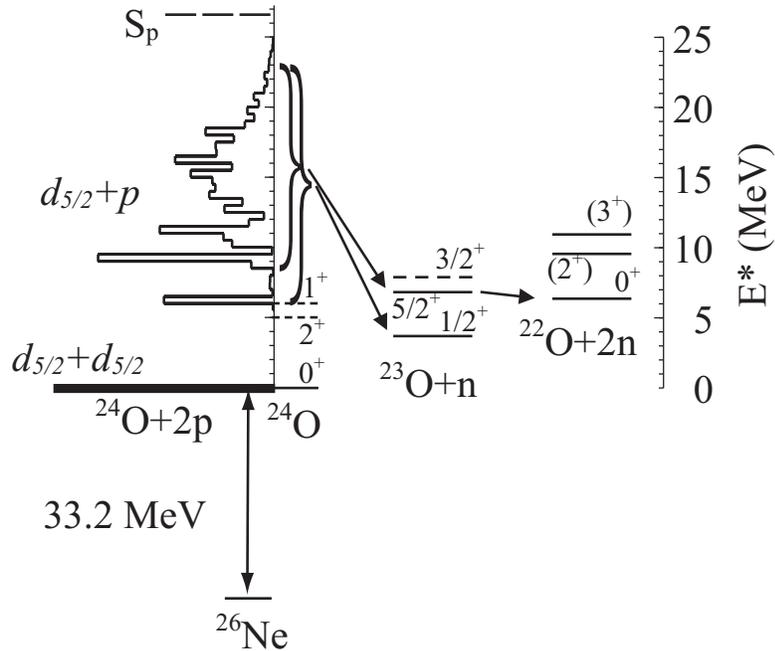,width=4.in}
\end{center}
\caption{Population of states in neutron-rich oxygen isotopes following the two-proton knockout reaction from $^{26}$Ne.}
\label{fig3}
\end{figure}

The population of the $5/2^+$ state in $^{23}$O is the result of a different two-proton knockout reaction. Instead of the knockout of the two $\pi(0d_{5/2})^2$ protons, the knockout involves one of the core $\pi(0p)^6$ protons in addition to one of the valence $\pi(0d_{5/2})^2$ protons leaving the system in a neutron-unbound excited state. The corresponding one proton (inner shell) knockout from various fluorine isotopes was used to explain the cross section of the population of neutron-rich oxygen isotopes \cite{Tho03}. 

Figure \ref{fig3} shows the excitation energy spectrum of this knockout scenario and the subsequent decay to the excited state of $^{23}$O, whose neutron decay to $^{22}$O is then detected in the present experiment. 

The selective population of the $5/2^+$ is due to the mixing of the excited proton configuration with neutron excitations. The proton configuration ($\pi(0p)\times\pi(0d_{5/2})^{-1}$) can mix with the neutron excitations of the form $\nu(0p)^{-1}\times\nu(0d_{3/2})^1$,
$\nu(1s_{1/2})^{-1}\times\nu(0f1p)^1$, and 
$\nu(0d_{5/2})^{-1}\times\nu(0f1p)^1$. The first of these has large
spectroscopic overlap with high-lying negative-parity excitations in $^{23}$O; the other two will excite one neutron
(either a $1s_{1/2}$ or a $0d_{5/2}$) to the continuum of the {\it fp}-shell.
The current setup is not geometrically efficient for the detection of these high-energy neutrons which probably contribute to the non-resonant background. The resulting $^{23}$O is then in the $1/2^+$ ground state or the $5/2^+$ excited state. The subsequent neutron decay of the $5/2^+$ state is the resonance observed in the present experiment. These different scenarios and the possible direct three-particle 2$p$1$n$ knockout is further described in References \cite{Fra07a,Sch07}.

In this reaction mechanism the $0d_{3/2}$ configuration is not populated and thus the recently observed $3/2^+$ (particle) state \cite{Ele07} is not populated in our experiment. The experiments are complementary because the present knockout reactions are sensitive to hole states while the single-neutron transfer reaction $^{22}$O(d,p)$^{23}$O$^*$ of Reference \cite{Ele07} is predominantly sensitive to particle states.

\section{Summary}

A two-proton knockout reaction was successfully used to populate neutron-unbound states in $^{23}$O. In contrast to the population of bound excited states that are predominantly populated from the knock-out of valence protons, the population of neutron-unbound states involves the knock-out of inner shell protons. The mixing of these deep-hole proton states with neutron states leads to the emission of a continuum neutron. The non-observation of any excited states in $^{24}$O and the 3/2$^+$ particle state in $^{23}$O and the observation of the 5/2$^+$ hole state in $^{23}$O are consistent with this reaction mechanism.

This new reaction mechanism offers the opportunity to explore excited states in neutron-rich nuclei that might not be accessible with any other method.

\section*{Acknowledgments}
We would like to thank the members of the MoNA collaboration G. Christian, C. Hoffman, K.L. Jones, K.W. Kemper, P. Pancella, G. Peaslee, W. Rogers, S. Tabor, and about 50 undergraduate students for their contributions to this work. We would like to thank R.A. Kryger, C. Simenel, J.R. Terry, and K. Yoneda for their valuable help during the experiment. Financial support from the National Science Foundation under grant numbers PHY-01-102533, PHY-03-54920, PHY-05-55366 PHY-05-55445, and PHY-06-06007 is gratefully acknowledged. J.E.F. acknowledges support from the Research Excellence Fund of Michigan.

\end{document}